\documentclass[conference]{IEEEtran}
\IEEEoverridecommandlockouts
\usepackage{cite}
\usepackage{amsmath,amssymb,amsfonts}
\usepackage{algorithmic}
\usepackage{graphicx}
\usepackage{textcomp}
\usepackage{xcolor}
\usepackage{svg} 
\def\BibTeX{{\rm B\kern-.05em{\sc i\kern-.025em b}\kern-.08em
    T\kern-.1667em\lower.7ex\hbox{E}\kern-.125emX}}
\begin{document}

\title{Increasing Line Outage Localization Performance with Ensemble Classifiers\\
}

\author{\IEEEauthorblockN{Daniel Flores and Yuanrui Sang}
\IEEEauthorblockA{\textit{Dept. of Electrical and Computer Engineering} \\
\textit{University of Massachusetts Amherst}\\
Amherst, MA, United States \\
\{dflores, ysang\}@umass.edu}
\and
\IEEEauthorblockN{Michael P. McGarry}
\IEEEauthorblockA{\textit{Dept. of Electrical and Computer Engineering} \\
\textit{University of Texas at El Paso}\\El Paso, TX, United States \\
mpmcgarry@utep.edu}
}

\maketitle

\begin{abstract}
In many cases, the outage of one transmission line in a system can be localized by monitoring the power flow of another line, and machine learning methods can be used to distinguish the cases under uncertainty. In this study, we examine the improvements in line outage localization performance achieved by various ensemble classifiers compared to single-model methods. In the case studies, we compared the classification results with measurement data collected at observed transmission lines (OTLs) selected using three algorithms, i.e, greedy maximum coverage problem (MCP), high-$\eta$, and random selection, based on two sensitivity factors, i.e., line outage distribution factors (LODFs) and line outage impact factors (LOIFs). We found that the OTLs selected by the greedy MCP algorithm yielded the highest $F_{1}$ score and the ensemble classifiers significantly outperformed a base kNN classifier. The extra-trees bagging technique achieved the highest $F_{1}$ score in many instances. All the findings were statistically significant.
\end{abstract}


\section{Introduction}
Transmission line outages are disturbances in a power system that can disrupt reliable operation and are triggered by events such as extreme weather, equipment malfunctions, human error, and cyberattacks \cite{ERN2021},\cite{LB2023}. Under stressed conditions, a single outage can potentially lead to cascading failures and system blackouts \cite{INKA2023}. As a result, accurately locating an outage is a critical first step in determining the best course of action for lessening its effects.

Finding the location of an outage depends heavily on the observability of the power system and on whether the available measurements can accurately distinguish different outage scenarios. The use of synchrophasor technology, such as phasor measurement units (PMUs), has been widely adopted in many studies for its ability to enable wide-area monitoring across the system. PMUs offer various technical advantages, including time-synchronized measurements at rates of 30-60 measurements per second, which traditional monitoring systems, such as Supervisory Control and Data Acquisition (SCADA) systems, lack \cite{SKJJ2020}. Because of this, PMUs are useful tools that can improve situational awareness across a power system and detect events such as outages in transmission lines and generators \cite{JCAT2022}. However, the deployment of PMUs comes with its own set of challenges, with cost being the main concern, as installing one at every bus in a system becomes impractical. Several studies have been conducted to address this by developing methods to optimally select locations within a system that maximize observability while minimizing the number of PMUs required \cite{JM2021} - \cite{ASG2021}.

Researchers have developed a wide range of techniques for PMU-based line outage detection and localization. Current methods can take analytical or a machine-learning approaches. Analytical methods rely on physics-based modeling or linear formulations of system behavior to detect and locate outages. One example of an analytical approach is presented in \cite{DBS2019}, where PMU measurements are used to detect outages through frequency disturbances and then locate them by analyzing changes in power flow. Similar analytical methods were presented in \cite{AKTB2019} and \cite{AKTB2020}, in which PMU measurements collected during a line outage were compared with offline simulated outage scenarios, selecting the scenario with the minimum least-squares error as the outage line. In the studies mentioned, it is assumed that PMU data is already available, and the proposed methods do not address PMU placement or measurement selection. 
In \cite{YA2022}, a Least Absolute Shrinkage and Selection Operator (LASSO)-based sparse estimation technique was used for outage localization with a limited number of PMUs selected through trial and error. Later, \cite{YA2025} shifted focus to PMU placement while still using LASSO, with the objective of minimizing the number of PMUs while reducing collinearity in the coefficient matrix for event detection, thereby improving distinguishability between different outage scenarios.
Machine learning approaches for line outage localization have gained attention, with supervised models being used for their ability to classify outage scenarios from labeled training data. The performance of models heavily depends on the selection of features. In \cite{ATC2019}, PMU placement and feature selection for line outage localization were addressed using power-grid partitioning and mutual information.
 
Line outage localization is a critically important capability of autonomous power system management. In prior work~\cite{FMY1025}, we introduced and analyzed
the performance of methods to localize line outages using machine learning. In that work, we introduced the Line Outage Impact Factor (LOIF) as a sensitivity factor to determine where to deploy monitoring devices, such as phasor measurement units (PMUs), to collect power measurements that are used as features for classification to localize line outages. LOIF is used to generate sets of line outages that could be detected using power measurements from a particular line in the power system; we call each of those lines an observed transmission line (OTL). Selecting OTLs then becomes a maximum coverage problem (MCP), in which we select the minimum number of OTLs with sets of detectable line outages that collectively cover all line outages. We will call this line outage coverage state  full coverage (FC). Then we used k-nearest neighbors (kNN), a single-model method, to localize the line outages and achieved reasonable results.

Ensemble classifiers combine the output of multiple base classifiers to produce a final classification result. Bagging ensembles, that average across parallel base classifiers, improve performance by reducing variance whereby boosting ensembles, that correct mistakes of prior classifiers stacked serially, improve performance by reducing bias and variance~\cite{HTF09}. In this work, we extend our prior work by uncovering the impact that ensemble classifiers have on line outage localization performance, as measured by $F_{1}$ score.

The contributions of the work is listed as follows:
\begin{itemize}
    \item This study proposes using ensemble methods for line outage localization, and carried out case studies on two test systems, the UIUC 150-bus system and the ACTIVSg200 system. Three ensemble methods, namely, AdaBoost, extra trees, and random forest, are compared to a single-model method, kNN, in terms of $F_1$ score of line outage localization based on data collected at the same OTLs, and results show that ensemble methods perform noticeably better than kNN.
    \item Classification results based on data collected at OTLs selected using three different methods, greedy maximum coverage problem (MCP), high-$\eta$, and random selection, are compared, and results show that greedy MCP yielded the best results overall.
    \item Classification results based on data collected at OTLs selected with two sensitivity factors, LOIF and line outage distribution factors (LODF) are compared, and results show that LOIF tends to outperform LODF in terms of OTL selection, especially in cases with a small number of OTLs.
\end{itemize}

In Section \ref{sec:otl}, we review the sensitivity factors (including LOIF) and algorithms that we use to select OTLs. In Section \ref{sec:classifiers}, we review classification using classifier ensembles outlining a few key techniques. In Section \ref{sec:plan}, we outline the experimental plan to analyze the performance improvements provided by ensemble classifiers, and then we present the results of the experiments in Section \ref{sec:results}. We conclude the article with a summary of our discoveries and paths for future investigation in Section \ref{sec:conclusion}.

\section{Selecting Observed Transmission Lines (OTLs)}
\label{sec:otl}
Selecting OTLs to provide power measurement data for line outage localization consists of three steps \cite{FYM0526}:
\begin{enumerate}
    \item Measuring a line's sensitivity to line outages on other lines,
    \item determining discernibility of line outages via power measurements from a candidate OTL, and
    \item selecting the set of OTLs.
\end{enumerate}

\subsection{Line Outage Sensitivity Factors}
\label{sec:sensitivity}
In this work the goal is to find a set of OTLs whose power flow measurements, captured at its connecting buses, provide discernibility between different single-line outage scenarios. During an outage there is a change in power flow on the OTL $a$, $\Delta P_a$, due to the outage of line $b$. These changes in power flow can be observed with the following sensitivity factors developed using the DC power flow model:
\begin{itemize}
    \item \textbf{Line Outage Distribution Factors (LODF):}\\
        Measures the portion of power flow from an outage line $b$ that will be redistributed on line $a$. This can be calculated by normalizing the change in power flow $\Delta P_a$ with the pre-outage power flow of line $b$, $P^{pre}_b$ as shown in Eq. \ref{eq:LODF}.
    \item \textbf{Line Outage Impact Factors (LOIF):}\\
        Measures the change in power flow on the OTL $a$ when line $b$ goes out. This can be calculated by normalizing the change in power flow $\Delta P_a$ with the pre-outage power flow of OTL $a$, $P^{pre}_a$ as shown in Eq. \ref{eq:LOIF}.
\end{itemize}
\begin{equation}
L_{a,b} = \frac{\Delta P_{a}}{P_b^{\text{pre}}},
\label{eq:LODF}
\end{equation}
\begin{equation}
\mathcal{O}_{a,b} = \frac{\Delta P_{a}}{P_a^{\text{pre}}},
\label{eq:LOIF}
\end{equation}

LOIF and LODF measure the same behavior but at different locations in the power system. LODF requires knowing the pre-outage power flow of the line that went out, which can be challenging if the power flow measurements are collected at a fixed set of OTLs, and measurements on the outage line are not present. In addition, it is assumed that the outage line has not yet been identified. LOIF on the other hand works well with the data collected at the OTLs, it can be used to calculate the expected post-outage power flow at the OTL $a$, $P^{post}_a$ due to the outage of line $b$
\begin{equation}
P_a^{\text{post}} = P_a^{\text{pre}} (1 + \mathcal{O}_{a,b}).
\end{equation}

\subsection{Set of Discernible Outages}
The objective of the OTL selection process is to determine a subset of transmission lines whose power-flow measurements provide sufficient information to distinguish among different single-line outage scenarios. Using LOIF and LODF introduced in Section \ref{sec:sensitivity}, outage scenarios can be evaluated based on how uniquely they affect the power flow measurements collected at candidate OTLs. An effective OTL should exhibit observable and distinguishable changes in power flow across different outages, allowing classification models to identify the outage condition accurately.

With LODF and LOIF, we can formulate an OTL selection method, where we determine which lines to collect power flow data. We first consider each transmission line in a system as potential OTLs, and determine which outages would be discernible based on the measurements collected at the OTL. This is done through a two-stage thresholding process where we:
\begin{itemize}
    \item Keep factors whose magnitude is greater than or equal to $\beta$. This removes outages whose impact on the OTL's power flow is low, which can be confused for power flow measurements of the system under normal conditions (no outage).
    \begin{equation}
c_1(a,b) = \left| \mathcal{O}_{a,b} \right| \geq \beta \mbox{.}
\end{equation}
    \item Remove factors with similar values by ensuring that the minimum difference between all LOIF/LODF values is greater than or equal to $\gamma$. This removes outages similar to each other, keeping outages with distinct impacts on the power flow of the OTL.
    \begin{equation}
c_2(a,b) = \forall_{c} \in \mathcal{L}\setminus \{a, b\} , \left| \mathcal{O}_{a,b} - \mathcal{O}_{a,c} \right|  \ge \gamma \mbox{.}
\end{equation}
\end{itemize}
By combining both threshold conditions, we obtain a final subset of outage lines for each possible OTL, $S_a$.
\begin{equation}
S_{a}(\beta,\gamma) = \{ \forall_{b} \in \mathcal{L} \;\Big|\; c_1(a,b) \land c_2(a,b)\} \mbox{.}
\end{equation}
For this work, we fix $\beta$ and $\gamma$ to 0.1 to find discernible outages for each OTL.

\subsection{OTL Selection Algorithms}
After each transmission line is evaluated as potential OTLs, the following selection algorithms are used to determine a set of lines from which we are interested in collecting power flow measurements.
\begin{itemize}
    \item \textbf{Coverage Metric (High $\eta$):} This metric evaluates the proportion of discernible outage scenarios associated with each candidate OTL over all transmission lines in the system. For OTL $a$, the coverage metric $\eta_a$ is defined as the ratio between the number of discernible outages in $S_a$ and the total number of lines in the system, $\mathcal{L}$, excluding the OTL itself, as shown in Eq. 7.
    
    \begin{equation}
    \eta_a = \frac{|S_a|}{|L|-1}
    \end{equation}
    
     A higher value of $\eta_a$ indicates that the OTL can distinguish among a greater number of outage scenarios, making it a more informative measurement location for outage classification. For a selection of $x$ OTLs, the candidate lines are ranked according to $\eta_a$ and the top OTLs are selected.
    \item \textbf{Maximum Coverage Problem (MCP)}: Although OTLs with high coverage can individually distinguish many outage scenarios, multiple OTLs may provide redundant outage information. To address this, the MCP formulation is used to select a subset of OTLs whose combined $S_a$ sets maximize the total number of covered outage scenarios. A greedy approach is implemented, where OTLs are iteratively selected to maximize additional outage coverage. Each newly selected OTL is evaluated based on the outage scenarios it contributes beyond those already covered by the previously selected OTLs. This process is continued until $x$ OTLs are selected or until there are no more uncovered outages.
\end{itemize}

\section{Ensemble Classifiers}
\label{sec:classifiers}
Classification, a branch of supervised machine learning, maps a set of data features to a nominal class label. In this work, the \textit{features} are active and reactive power at the from bus of the selected OTLs (PF and QF). The \textit{class label} is the outage line number; a 0 indicates no outage. Once trained, a classification model can predict the class label of unseen data samples based on the patterns learned from the training data. For example, the k-Nearest Neighbors (kNN) classification model stores labeled training data and determines the label of a test sample by calculating distances to neighboring samples and selecting the majority label among its $k$ closest neighbors.

Ensemble classifiers perform the classification task using two or more classifiers trained independently and combine their results for a final classification output. Two popular ensemble classifier methods are boosting and bagging.


\subsection{Boosting}
\begin{figure}[ht!]
    \centering
    \includegraphics[width=0.8\linewidth]{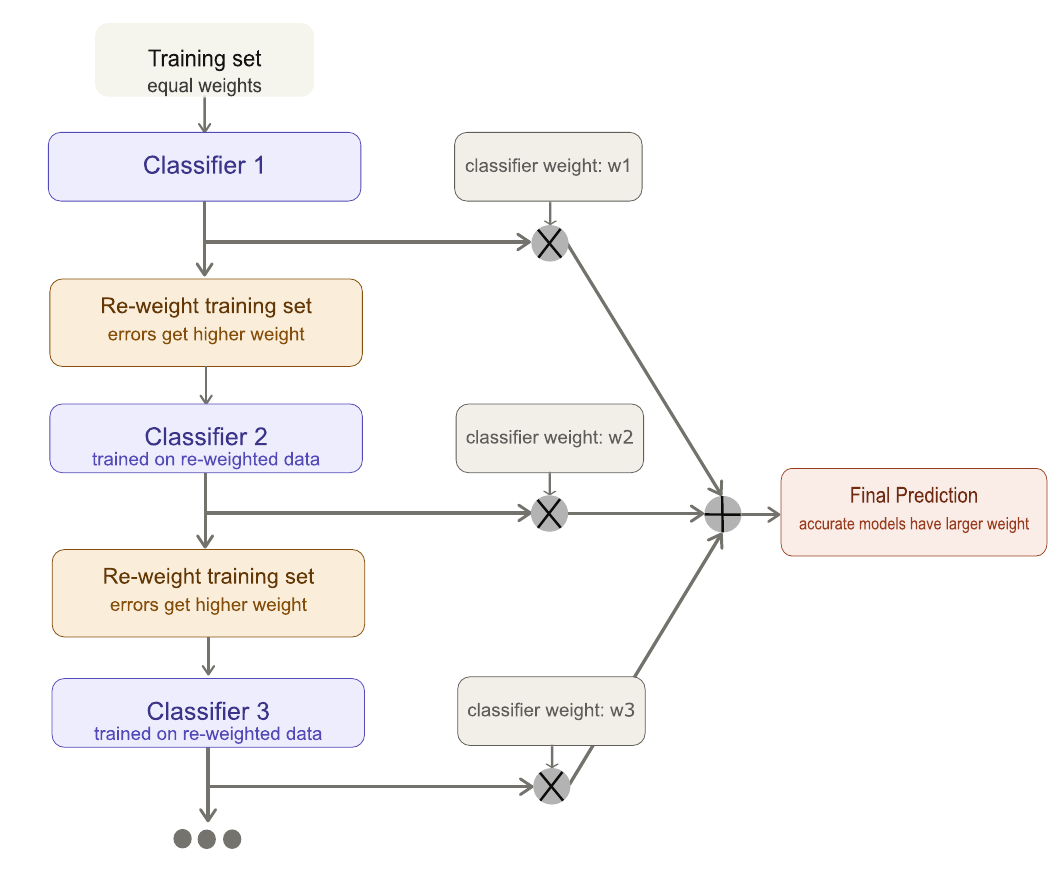}
    \caption{Boosting ensembles of classifiers train multiple classifiers in series whereby subsequent classifiers are trained to correct the prediction errors of the previous classifier. Classifier outputs are combined using a weighted average with more accurate classifiers having a larger weight.}
    \label{fig:boosting}
\end{figure}

Boosting ensembles of classifiers train multiple classifiers in series whereby subsequent classifiers focus on correct classification of the training samples that were incorrectly predicted by the prior classifiers. Classifier outputs are combined using a weighted sum; classifiers that were more accurate have a larger weight in the weighted sum; see Figure \ref{fig:boosting}. Classification performance is increased by reducing bias. The case studies in this paper utilize one boosting method, the Adaptive Boosting with Decision Trees, described below. 
\subsubsection*{Adaptive Boosting with Decision Trees}
Adaptive Boosting (AdaBoost) with Decision Trees is an ensemble classifier model that combines multiple decision trees. When training the model, decision trees (DT) are used sequentially, with each tree trained on the same dataset and assigning greater weight to misclassified labels from the previous DT. This process is repeated until a specified number of trees is reached or until the model achieves a low classification error.

\subsection{Bagging}
\begin{figure}[ht!]
    \centering
    \includegraphics[width=0.8\linewidth]{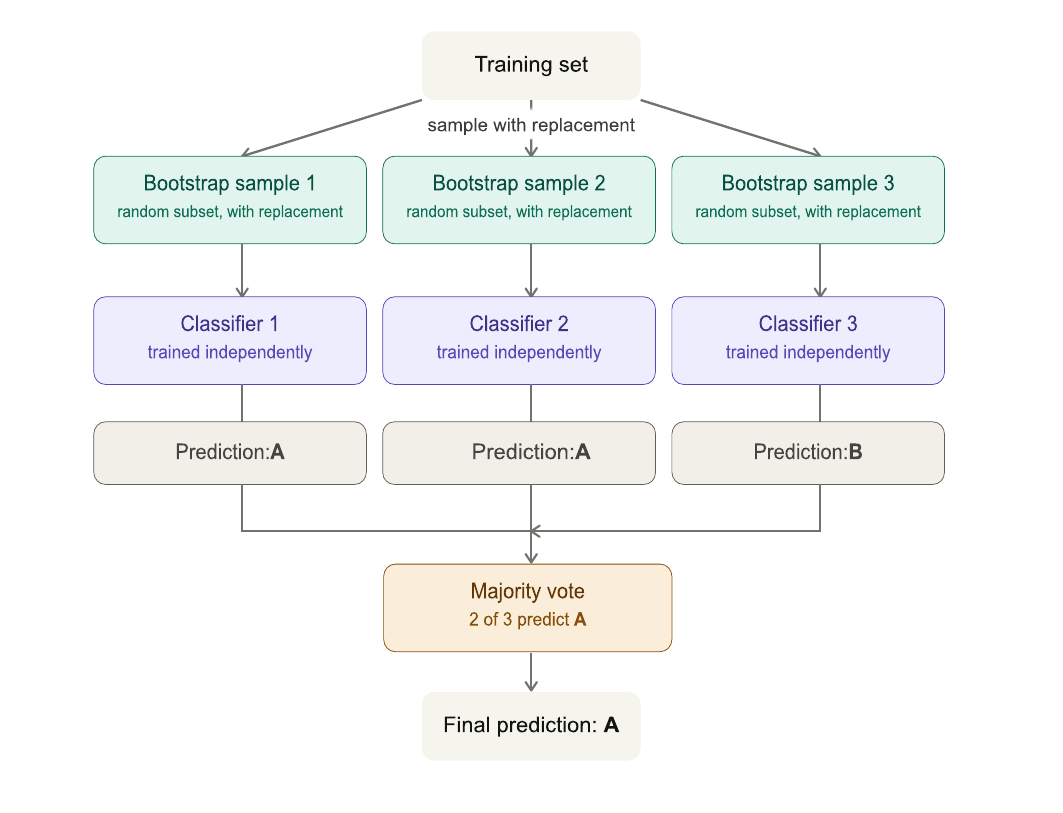}
    \caption{Bagging ensembles of classifiers train multiple classifiers in parallel using different samples and features from the training data set. Classifier outputs are either averaged to produce the combined classifier output or a majority vote is used.}
    \label{fig:bagging}
\end{figure}

Bagging ensembles of classifiers train multiple classifiers in parallel using different subsets of the training data set (subsets of features and samples). Classifier outputs are combined by averaging or by a majority vote; see Figure \ref{fig:bagging}. Classification performance is increased by reducing variance. The case studies in this paper utilize two bagging methods, Random Forest and Extra Trees, described below.

\subsubsection*{Random Forest}
The Random Forest (RF) classifier is similar to the AdaBoost classifier in that it uses multiple decision trees (DTs) to perform classification. However, RF trains each DT independently using a random bootstrap sample of the training data and a random subset of features at each split. Adding randomness allows diversity among the DTs, meaning that the errors are decoupled from one another. This decreases the variance of the ensemble model by combining the prediction probabilities of multiple diverse DTs and obtaining an average probability for each label. In addition, each DT uses a best-split approach by testing multiple thresholds across randomly selected features and selecting the split that minimizes impurity.
\subsubsection*{Extra Trees}
Extra Trees (ET) are similar to the RF classifier but use additional randomness in each individual DT model. Instead of finding the best possible thresholds, ET randomly selects a threshold for each split, often reducing training time.

\section{Experimental Plan}
\label{sec:plan}
\subsection{Objective}
A set of experiments were conducted to compare the classification performance between the two proposed OTL selection algorithms and a random selection of OTLs. In order to remove bias from the randomly selected lines, we get an aggregate average $F_1$ score from ten sets of randomly selected OTLs. In order to train and evaluate the performance of our classification models, we collected labeled datasets generated with \textsf{\textsc{Matpower}} simulations. These labeled datasets are class-balanced and were generated using stratified sampling in which bus loads experienced random load perturbations between $\pm5\%$. In this sampling, each transmission line was disconnected to simulate a single-line outage where the following procedure is implemented:
\begin{enumerate}
    \item Run AC Optimal Power Flow (OPF) Solution of the system under normal conditions (pre-outage) with random load conditions and collect generation data.
    \item Under the pre-outage conditions for load and generation, disconnect the line and run AC Power Flow (PF) Solution to observe the power flow redistribution.
    \item Check for convergence for both OPF and PF solutions. If there is no convergence for the given line disconnection or for the random load conditions, do not collect power flow measurements at the OTLs. If the solutions converge, collect the power flow measurements and provide a label for the given line disconnection (number of the line).
    \item Repeat steps 1-3 for each line in the system.
\end{enumerate}
\subsection{Experiments}
Classification experiments were conducted and designed to vary the following parameters:
\begin{itemize}
    \item Sensitivity Factor: [LODF, LOIF]
    \item OTL Selection Algorithm: [Greedy MCP, High $\eta$, Random]
    \item $x$ - Number of OTLs selected: [1, 2, 4, 8, Full Coverage]
    \item Classification Model: [kNN, AdaBoost, RF, ET]
\end{itemize}
A total of 120 individual classification experiments were repeated 24 times, each with a new labeled dataset. The Python package \texttt{scikit-learn} was used to implement the classifiers. Average $F_1$ score was used to evaluate performance. The classification experiments were conducted for two power systems:
\begin{itemize}
    \item UIUC 150-Bus System \cite{UIUC}
    \item ACTIVSg 200-Bus System \cite{ACTIVSg}
\end{itemize}

\section{Experimental Results}
\label{sec:results}
\begin{figure*}[ht!]
    \centering
    \includegraphics[width=0.85\textwidth]{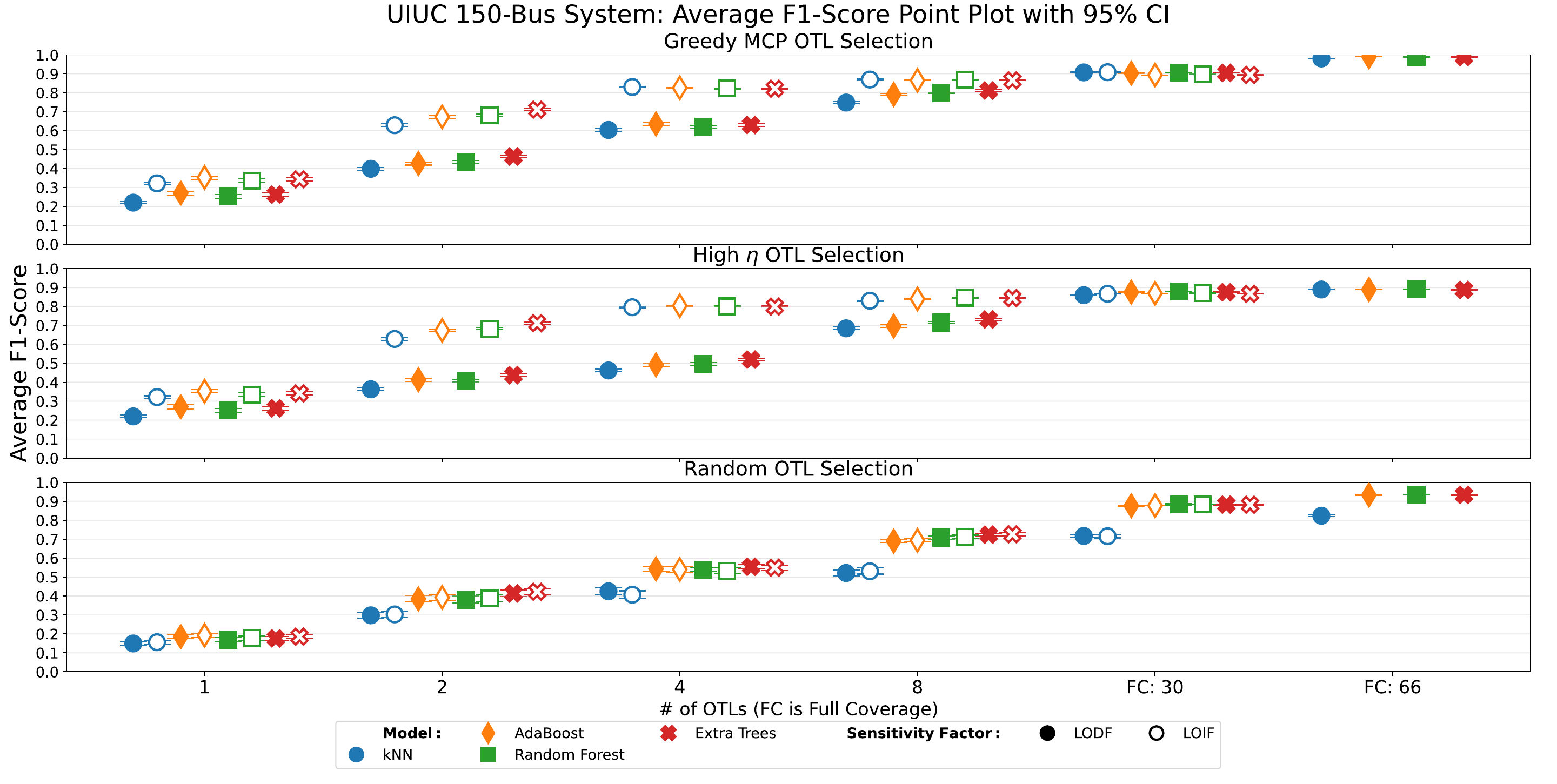}
    \caption{Average $F_1$ scores for line outage localization in the UIUC 150-bus system using different sensitivity factors, OTL selection algorithms, and classification models across varying numbers of selected OTLs.}
    \label{fig:uiuc}
\end{figure*}

\subsection{150-Bus System (UIUC)}
In the 150-bus system, 190 of the 217 transmission lines converged on their power flow (PF) solutions when simulating their outages. With greedy MCP, 30 OTLs covered all outages using LOIF, while 66 were needed with LODF. For 1-8 OTLs, Figure \ref{fig:uiuc} shows that LOIF consistently outperformed LODF for both the greedy MCP and high-$\eta$ methods. For 1 OTL, the best performance was achieved by LOIF, with $F_1$ scores around 0.32-0.35, while ensemble models had a slight edge over kNN, with a small difference of 0.03 in the $F_1$ score. With LODF, the performance was lower, achieving $F_1$ scores around 0.22 - 0.27 with an average difference of 0.09 compared to LOIF. 
For 2 OTLs, the performance was the same for both the greedy MCP and high-$\eta$ methods, with $F_1$ scores around 0.61 - 0.71 for LOIF and 0.40 - 0.46 for LODF. ETs achieved the highest scores, and kNN the lowest. 
By 4 OTLs, LOIF still performs better than LODF, but the $F_1$ scores stay relatively the same across all classification models, with $F_1$ scores of 0.83 and 0.80 for the greedy MCP and high-$\eta$ methods, respectively.
For 8 OTLs, the same trend was observed with LOIF, with $F_1$ scores increasing slightly by 0.87 and 0.83 for greedy MCP and high eta, respectively. 
For the LODF-based cases, the $F_1$ score were similar for results obtained with OTLs selected using the high-$\eta$ method and random selection, showing that LODF has limited capability of effectively indicating whether a line can enhance system observability.
Under full coverage with 30 OTLs, LOIF and LODF performed the same across all models, but greedy MCP performs better, achieving average $F_1$ Scores of 0.90 and 0.87 for high eta. Compared to the 8 OTLs, there is only a small improvement in $F_1$ scores, indicating that most outages can already be correctly classified using a relatively small number of strategically selected transmission lines. With 66 OTLs, we get near-perfect $F_1$ scores with LODF but need twice as many lines as with LOIF, which can result in higher monitoring and communication costs.
\subsection{200-Bus System (ACTIVSg200)}
\begin{figure*}[ht!]
    \centering
    \includegraphics[width=0.85\textwidth]{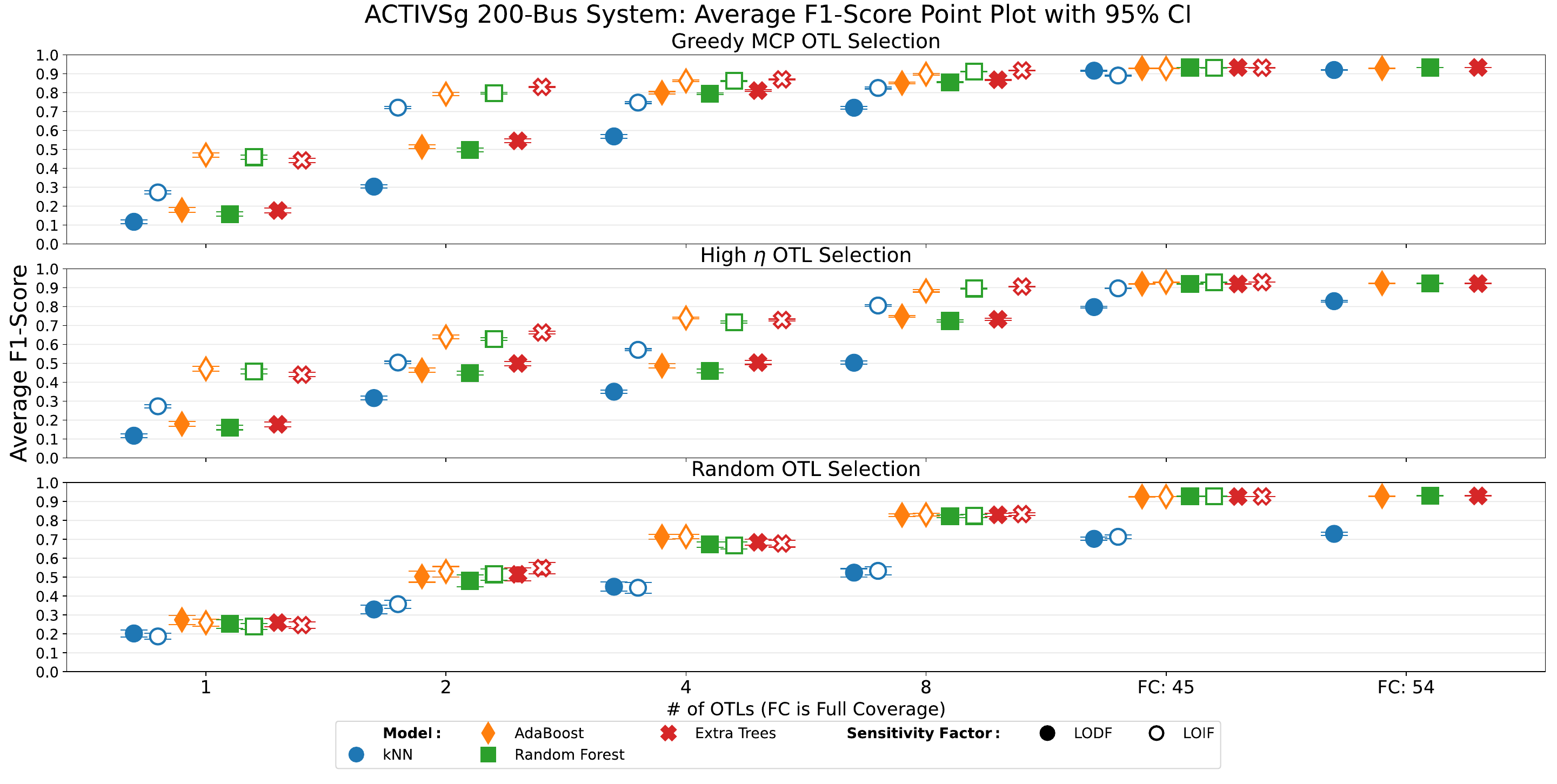}
    \caption{Average $F_1$ scores for line outage localization in the ACTIVSg200 system using different sensitivity factors, OTL selection algorithms, and classification models across varying numbers of selected OTLs.}
    \label{fig:ACTIVSg200}
\end{figure*}

For the 200-bus system, 184 out of 245 transmission lines converged when simulating their outages. Under greedy MCP, we achieved full coverage for all outages with only 45 OTLs for LOIF, while LODF required 54 OTLs. As shown in the Fig. \ref{fig:ACTIVSg200}, ensemble models achieved better performance across different numbers of selected OTLs than the distance-based classification model, kNN. Among the selection algorithms, greedy MCP appeared to select better sets of lines than both the high-$\eta$ method and random selection across all numbers of selected OTLs. For 1 OTL, the best performance was achieved with LOIF, with $F_1$ scores around 0.44-0.46 for ensemble models, while kNN performed the worst with an $F_1$ score of 0.27. When using 2 OTLs, the ET model performed best under the greedy MCP algorithm, with maximum $F_1$ scores of 0.83 and 0.55 for LOIF and LODF, respectively. Compared to greedy MCP, the high-$\eta$ method achieved maximum $F_1$ scores of only 0.66 and 0.50 for LOIF and LODF, respectively, using the RF model. This trend continues as we increase the number of OTLs to 4, where LOIF selects better OTLs and all ensemble models achieve $F_1$ scores around 0.86 under greedy MCP. For LODF, the $F_1$ score was around 0.80. When using the High-$\eta$ algorithm, all $F_1$ scores were below 0.75 for LOIF and below 0.50 for LODF. For 8 OTLs, we began to achieve $F_1$ scores close to 0.90 for the ensemble models when using LOIF for both the greedy MCP and high-$\eta$ methods. 
Under full coverage for both LOIF and LODF, the $F_1$ score remains nearly the same across all ensemble models and selection algorithms, with $F_1$ scores staying around 0.92-0.93. Compared to 8 OTLs, there is only a minimal change in $F_1$ scores, indicating that a majority of outages can be correctly classified with a small number of lines, resulting in only 3\% of all lines in the system being used for monitoring purposes.
Similar to the 150-bus system, we noticed that the randomly selected OTLs yielded similar classification results as the high-$\eta$ method-selected line when the selection was based on LODF, confirming that LOIF is a better indicator of whether a line is a good OTL.

\section{Conclusion}
\label{sec:conclusion}
In conclusion, using sensitivity factors such as LOIF enables better selection of OTLs within a power system for monitoring single-line outages than LODF. Maximizing outage coverage with greedy MCP, rather than selecting OTLs with the highest coverage metric, results in better classification performance across all models. In addition, ensemble classifiers consistently outperformed the baseline kNN classifier across both power systems. Although near-perfect $F_1$ scores could be achieved under full coverage, the results show that most outage scenarios can be accurately localized using only a small number of strategically selected OTLs, thereby reducing monitoring requirements and associated costs for the practical deployment of PMUs. In future work, we plan to partition the power system into regions to enhance the scalability of the proposed line outage localization method.


\begin{thebibliography}{00}
\bibitem{ERN2021} S. Ekisheva, R. Rieder, J. Norris, M. Lauby, and I. Dobson, “Impact of Extreme Weather on North American Transmission System Outages,” in 2021 IEEE Power \& Energy Society General Meeting (PESGM), Jul. 2021, pp. 01–05.
\bibitem{LB2023} L. Bader et al., “Comprehensively Analyzing the Impact of Cyberattacks on Power Grids,” in 2023 IEEE 8th European Symposium on Security and Privacy (EuroS\&P), Jul. 2023, pp. 1065–1081.
\bibitem{INKA2023} S. Imai, D. Novosel, D. Karlsson, and A. Apostolov, “Unexpected Consequences: Global Blackout Experiences and Preventive Solutions,” IEEE Power and Energy Magazine, vol. 21, no. 3, pp. 16–29, May 2023.
\bibitem{SKJJ2020} H. R. Sudheendra, V. Kumar Jadoun, N. S. Jayalakskmi, and A. Agarwal, “Latest Trends in PMU Placement Techniques,” in 2020 International Conference on Power Electronics \& IoT Applications in Renewable Energy and its Control (PARC), Feb. 2020, pp. 340–345.
\bibitem{JCAT2022} M. R. Jegarluei, J. S. Cortés, S. Azizi, and V. Terzija, “Wide-Area Event Identification in Power Systems: A Review of the State-of-the-Art,” in 2022 International Conference on Smart Grid Synchronized Measurements and Analytics (SGSMA), May 2022, pp. 1–7.
\bibitem{JM2021} T. Johnson and T. Moger, “A critical review of methods for optimal placement of phasor measurement units,” International Transactions on Electrical Energy Systems, vol. 31, no. 3, p. e12698, 2021.
\bibitem{RSK2021} V. V. R. Raju, K. H. Phani Shree, and S. V. Jayarama Kumar, “Measurement Redundancy constrained Optimal PMU Locations,” in 2021 International Conference on Sustainable Energy and Future Electric Transportation (SEFET), Jan. 2021, pp. 1–6.
\bibitem{ZSIZ2025} M. Zekry, M. M. Samy, D. K. Ibrahim, and E. M. Aboul-Zahab, “Optimal PMUs Placement for Enhanced Power System Observability,” in 2025 26th International Middle East Power Systems Conference (MEPCON), Dec. 2025, pp. 1–7.
\bibitem{ASG2021} Mohd. N. Ansari, R. K. Singh, and A. K. Gupta, “Optimal PMU Positioning Considering the Effect of Zero Injection Buses in Advanced Grid Monitoring,” in 2021 IEEE 2nd International Conference On Electrical Power and Energy Systems (ICEPES), Dec. 2021, pp. 1–5.
\bibitem{DBS2019} X. Deng, D. Bian, D. Shi, W. Yao, Z. Jiang, and Y. Liu, “Line Outage Detection and Localization via Synchrophasor Measurement,” in 2019 IEEE Innovative Smart Grid Technologies - Asia (ISGT Asia), May 2019, pp. 3373–3378.
\bibitem{AKTB2019} M. Alam, S. Kundu, S. S. Thakur, and S. Banerjee, “A new algorithm for single line outage estimation,” in 2019 Devices for Integrated Circuit (DevIC), Mar. 2019, pp. 113–117. 
\bibitem{AKTB2020} M. Alam, S. Kundu, S. S. Thakur, and S. Banerjee, “PMU based line outage identification using comparison of current phasor measurement technique,” International Journal of Electrical Power \& Energy Systems, vol. 115, p. 105501, Feb. 2020.
\bibitem{YA2022} T. Yildiz and A. Abur, “Improved Line Outage Detection in Transmission Systems with Few PMUs,” in 2022 North American Power Symposium (NAPS), Oct. 2022, pp. 1–6. 
\bibitem{YA2025} T. Yildiz and A. Abur, “Sparse PMU Placement Algorithm for Enhanced Detection and Identification of Power Grid Events,” IEEE Transactions on Power Systems, vol. 40, no. 2, pp. 1843–1853, Mar. 2025.
\bibitem{ATC2019} R. Alhalaseh, H. A. Tokel, S. Chakraborty, G. Alirezaei, and R. Mathar, “PMU Placement with Power Grid Partitioning for Line Outage Detection,” in 2019 4th International Conference on Smart and Sustainable Technologies (SpliTech), Jun. 2019, pp. 1–6.
\bibitem{FMY1025} D. Flores, M. P. McGarry, and Y. Sang, ``Effective Feature Selection for Line Outage Localization: Line Outage Impact Factors (LOIF),'' 2025 57th North American Power Symposium (NAPS), IEEE, Oct 2025.
\bibitem{HTF09} T. Hastie, R. Tibshirani, and J. Friedman, ``The Elements of Statistical Learning,'' 2009.
\bibitem{FYM0526} D. Flores, Y. Sang, and M. P. McGarry, “Line Outage Impact Factor (LOIF): A New Sensitivity Factor for Enhanced Transmission Observability,” Jun. 15, 2026, arXiv: arXiv:2606.17314. doi: 10.48550/arXiv.2606.17314.
\bibitem{UIUC} “UIUC 150-Bus System.” Accessed: Jun. 17, 2026. [Online]. Available: https://electricgrids.engr.tamu.edu/electric-grid-test-cases/uiuc-150-bus-system/
\bibitem{ACTIVSg} “Illinois 200-Bus System: ACTIVSg200.” Accessed: Jun. 17, 2026. [Online]. Available: https://electricgrids.engr.tamu.edu/electric-grid-test-cases/activsg200/



\end{thebibliography}
\end{document}